\documentclass[12pt,epsf,amssymb]{article} \textheight =23 truecm
\def\lsim{\raise0.3ex\hbox{$<$\kern-0.75em\raise-1.1ex\hbox{$\sim$}}}
\def\gsim{\raise0.3ex\hbox{$>$\kern-0.75em\raise-1.1ex\hbox{$\sim$}}}
\def\noi{\noindent} \def\nn{\nonumber} \def\bea{\begin{eqnarray}}
\def\eea{\end{eqnarray}} \def\beq{\begin{equation}}
\def\eeq{\end{equation}} 
\def\beeq{\begin{eqnarray}} \def\eeeq{\end{eqnarray}} \def\R{ {\rm R
\kern -.31cm I \kern .15cm}} \def\C{ {\rm C \kern -.15cm \vrule
width.5pt \kern .12cm}} \def\Z{ {\rm Z \kern -.27cm \angle \kern
.02cm}} \def\N{ {\rm N \kern -.26cm \vrule width.4pt \kern .10cm}}
\def\1{{\rm 1\mskip-4.5mu l} }

\usepackage{graphics} \usepackage{graphicx} \usepackage{epsfig}

\begin{document} 
\begin{center} 

{\large \bf Sum rules of Bjorken-Uraltsev type in the Bakamjian-Thomas relativistic quark model}

\par \vskip 10 truemm

 {\bf A. Le Yaouanc, L. Oliver and J.-C. Raynal}

\par \vskip 2 truemm

{\it Laboratoire de Physique Th\'eorique}\footnote{Unit\'e Mixte de
Recherche UMR 8627 - CNRS }\\    {\it Universit\'e de Paris XI,
B\^atiment 210, 91405 Orsay Cedex, France} 

\end{center}
\par \vskip 6 truemm

\begin{abstract}

The Bakamjian-Thomas relativistic quark model, describing hadrons with a fixed number of constituents, yields  in the heavy quark limit of QCD covariant Isgur-Wise functions and satisfies the whole tower of lowest moment sum rules (Bjorken-Uraltsev type sum rules). We first recall, as well as earlier results, the new formalism presented in our recent papers on Lorentz representations, which provide an elegant framework for the analysis of this model in the heavy quark limit and stress the results which have been already obtained in this direction. 
Then, we give some very explicit demonstrations of the fact that the Bakamjian-Thomas framework satifies the sum rules by considering simple cases of Isgur-Wise functions. In addition to the specific Bjorken and Uraltsev sum rules, an important  sum rule that involves only heavy mesons with light cloud $j^P = {1 \over 2}^-$ and their radial excitations is demonstrated. This latter sum rule is phenomenologically interesting because it constrains the derivatives of the radially excited Isgur-Wise functions at zero recoil. On the other hand, we recall the limitations of the Bakamjian-Thomas scheme. At finite mass, current matrix elements with the current coupled to the heavy quark are no longer covariant, and higher moment sum rules that hold in the heavy quark limit of QCD are not satisfied.

\vskip 10 truemm
\noi LPT-Orsay-15-75 \qquad\qquad October 2015 \par 
\vskip 8 truemm

\end{abstract}

\newpage
\section{Introduction} \hspace*{\parindent}
We give a rather extended introduction, in order to make understand the motivation of the calculations we are presenting.

\subsection{OPE sum rules in the heavy quark limit of QCD}

The heavy quark limit of QCD implies poweful constraints on form factors. In the elastic case for transitions ${1 \over 2}^- \to {1 \over 2}^-$ for the light cloud $\overline{B} \to D^{(*)} \ell \nu$, this limit implies that all form factors are given in terms of a single function, the famous Isgur-Wise (IW) function $\xi(w)$ \cite{IW-1}.\par
The OPE leads in general to sum rules, which take a simple form in the heavy quark limit.
A sum rule (SR) formulated by Bjorken \cite{BJORKEN} in the heavy quark limit of QCD implies the lower limit $\rho^2 = -\xi'(1) \geq {1 \over 4}$. This SR was formulated in a transparent way by Isgur and Wise, in terms of IW functions $\tau_{1/2}(w), \tau_{3/2}(w)$ for inelastic transitions ${1 \over 2}^- \to {1 \over 2}^+, {3 \over 2}^+$ at zero recoil  \cite{IW-2}.\par
Ten years later, a new SR was discovered by Uraltsev \cite{URALTSEV}, making use of the non-forward amplitude $\overline{B}(v_i) \to D^{(n)}(v') \to \overline{B}(v_f)$ ($v_i \not = v_f$). Uraltsev SR combined with Bjorken's yields the much more powerful lower bound for the elastic slope, $\rho^2 \geq {3 \over 4}$.\par

At this stage, it is important to make precise which type of sum rules we consider.
Indeed, we have to point out that we are dealing here with the lowest moment SR of a more general class of SR that also hold in the heavy quark limit of QCD. We do not consider SR that involve IW functions with powers of level spacings $\Delta E^{(n)}$, i.e. sums of the form $\sum_n (\Delta E^{(n)}_j)^{k} \mid \tau_j^{(n)}(w) \mid^2$ ($j = {1 \over 2},  {3 \over 2}$) where $k > 0$ is an integer. For $k = 1$ one has at zero recoil Voloshin SR for the HQET parameter $\overline{\Lambda}$ \cite{VOLOSHIN} and the counterpart for $k = 1$ of Uraltsev SR \cite{URALTSEV,LMMOPR}, while for $k = 2$ there are the SR at zero recoil for the important HQET parameters $\mu_G^2$ and $\mu_\pi^2$, formulated by I. Bigi et al. \cite{BIGI}. The general case for any value of $k$ has been formulated by Grozin and Korchemsky \cite{GK}. In the present paper we are concerned with the lowest moment case $k = 0$. 

In a number of papers we generalized Bjorken and Uraltsev SR, and we obtained a whole tower of SR that allow to constrain the higher derivatives of the elastic IW function $\xi(w)$. In particular, we found lower bounds on the successive derivatives \cite{LOR-1,LOR-2}, and also an improved lower bound of the curvature in terms of the slope \cite{LOR-3}. Similar results were also formulated for the baryon case $j^P = 0^+$ \cite{LOR-4}.\par

\subsection{Overview of our recent work}

Then, our research shifted to study a possible physical insight into these powerful results of the SR method. To this aim, we started from a general idea formulated very early by Falk \cite{FALK}, namely that the IW functions, leaving spin complications aside, originate in the possibility of a factorisation of the current matrix elements into a free heavy quark current and a part relative only to light quarks and which bears the QCD interaction, that within factorization is then a soft strong interaction. Then the IW functions correspond to the latter,  and have roughly the simplified form, e.g. for a scalar ground state :
\beq
\label{2.1e}
\xi(w) =\ < U(\Lambda_f)\varphi(v_0) \mid U(\Lambda_i)\varphi(v_0) >\ =\ < \varphi(v_0) \mid U(\Lambda)\varphi(v_0) > 
\eeq 

\noindent where $U(\Lambda)$ is a boost, the state $\mid \varphi(v_0) >$ with $v_0 = (1,0,0,0)$ represents the light cloud at rest (eigenstates of angular momentum), and the Lorentz transformation $\Lambda$ satisfies $\Lambda v_0 = v$ with $v^0 = w$.\par

An important hypothesis has been done in writing expression (\ref{2.1e}), namely the neglect of {\it hard gluon radiative corrections}. We comment shortly on this assumption at the end of the article.

These overlaps are covariant. With this ansatz of Falk, the light cloud belongs to a distinct Hilbert space and transforms itself according to a unitary representation of the Lorentz group. This corresponds to the covariance property of the overlap. One can then decompose these representations with the help of Lorentz group representation theory \cite{NAIMARK} in terms of irreducible representations of the Lorentz group, which must themselves be decomposed further into irreducible representations of $SU(2)$ to obtain definite $j$ states.

The decomposition of the  representation of the hadron light quark state leads to a parallel description of the IW function as an integral over "irreducible IW functions" {\it with a positive measure}. The integration is over a parameter $\rho$ introduced by Na\"{\i}mark \cite{NAIMARK}, which labels the irreducible unitary Lorentz group representations : 
\beq
\label{2.2e}
\xi(w) = \int \xi^\rho(w)\ d\nu(\rho)
\eeq
with $\xi^\rho(w)$ the irreducible IW functions and $d\nu(\rho)$ the positive measure. This integral representation in turn leads to a set of powerful bounds on the IW function. We have developped in detail the above Lorentz group analysis in two papers, that in the following we will quote as Lorentz I \cite{LOR-5} and Lorentz II \cite{LOR-6}. 

The formulation of IW functions as overlaps (i.e. scalar products) of light quark states also leads to a transparent presentation of sum rules, just using closure for products of such overlaps. 

In the case of ground state heavy baryons $j^P = 0^+$ or mesons $j^P = {1 \over 2}^-$ (for the light cloud), we have demonstrated that the constraints on the IW function that one has obtained from the sum rules of Bjorken and Uraltsev and their generalizations - which in certain cases require many steps -, can be derived quite directly from the above analysis of the elastic IW function by the Lorentz group method (\ref{2.2e}). In addition, many other bounds are found.

The integral representation (\ref{2.2e}) has been inverted in ref. \cite{LOR-6}, obtaining the measure in terms of the physical IW function $\xi(w)$ in the case  $j^P = {1 \over 2}^-$:
\beq
\label{2.3e}
{d\nu(\rho) \over  d\rho} = {1 \over 2\pi}\int_{-\infty}^{+\infty} e^{i\tau\rho}\ d\tau\ {1 \over 2\cosh\left({\tau \over 2}\right)} {d \over d\tau} \left[(\cosh(\tau)+1)\sinh(\tau)\xi(\cosh(\tau))\right]
\eeq
where $\tau$ is such that $w = \cosh(\tau)$. This integral should be {\it positive}.

One notes that the measure is given by a Fourier transform, where $\tau$ is related to the momentum transfer $w-1$, and $\rho$ is conjugate to it. It is then a sort of relativistic generalisation of the relation holding in the non relativistic case \cite{IWNR}, where the definite positive charge density in $\vec{r}$ is the Fourier transform of the form factor, which plays the role of $\xi(w)$, and that leads to similar bounds on the derivatives ot this form factor. 

The positivity condition on the r.h.s. of eqn. (\ref{2.3e}) has allowed us to test the consistency of quite a number of models given in the literature for the elastic meson IW function, checking the explicit formula of $\xi(w)$ for each of these models. We have considered a number of phenomenological formulas and also some quark models for $\xi(w)$.\par
Among the few quark models in the literature for the IW function that we have been able to examine, only two have passed these tests : the Bakamjian-Thomas (BT) relativistic quark model \cite{BT,TERENTEV,KEISTER,CARDARELLI,LOPR-1,GRS}, and the Bauer, Stech and Wirbel (BSW) model \cite{BSW}. As we emphasize below, we do not pretend to have been exaustive in this study of relativistic quark models for the elastic IW function.\par

Here, one can formulate an observation which may be useful for future analysis. For any {\it general approach} which formulates the matrix elements in terms of an arbitray set of rest frame wave functions, eigenstates of the generic spectroscopic mass operator, the above test on positivity of the inverted measure  (eqn.(\ref {2.3e})) should be satisfied for {\it any wave function} describing the ground state, or equivalently, the HQET constraints on the derivatives of the ground state IW function should be satisfied for any wave function.

We have then applied the same techniques to give a new formulation of the Bakamjian-Thomas relativistic quark model in its heavy quark limit, in Section 11 of Lorentz II \cite{LOR-6}. 

\subsection{The sum rules and the HQET constraints in the Bakamjian-Thomas relativistic quark model}

Let us indeed return to the sum rules and consider whether Bakamjian-Thomas relativistic quark models satisfy the HQET sum rules and the above constraints. 

First let us recall that previously, in our early studies, we have demonstrated in a rather general manner, in the form of quark-hadron duality, that the Bjorken and Uraltsev sum rules are satisfied \cite{LOPR-2,LOPR-3}.\par
We think now that a somewhat new demonstration can be presented, by passing through the Falk factorisation ansatz. Indeed, this postulate can be demonstrated very simply in the BT approach, in the original approach with Wigner rotations (see for example \cite{LOPR-1}. The overlaps are given by a very simple expression in terms of the internal wave functions for the light quarks, valid for any number of quarks, and the sum rules derive straightforwardly. It remains then to establish the covariance of these overlaps. 

On the other hand, we believe also that one can construct in full generality a manifestly covariant general version of these BT overlaps, which would then represents the BT model in the heavy quark limit in a simple covariant form. We have explicitly shown this point in Lorentz II \cite{LOR-6}, Section 11, for two simple cases : the elastic ground state overlap, and the one for transitions from the ground state to the $L=1, j=1/2, j=3/2$  orbital excitations. One recovers indeed our old covariant expressions for BT \cite{MORENAS-1,MORENAS-2}. 

Having represented the $\xi(w)$ of BT in the form of eqn. (\ref{2.1e}), this implies that it satisfies all the HQET constraints we have derived from it, and which are in accordance with what we have derived from the sum rules. Moreover, in Section 11 of Lorentz II \cite{LOR-6}, we give an explicit demonstration of the positivity of the r.h.s. of eqn.(\ref{2.3e}) in the BT class of models, which is a confirmation of the validity of these constraints in the model.

Now, in face of all the demonstrations, it is also advisable to ascertain the validity of the sum rules in BT by a direct and completely explicit calculation, using our early established expressions of IW functions for these states or their radial excitations, for the transitions ${1 \over 2}^- \to {1 \over 2}^-$ \cite{LOPR-1} and for the parity changing transitions ${1 \over 2}^- \to {1 \over 2}^+, {3 \over 2}^+$ \cite{MORENAS-1,MORENAS-2}.  

Therefore, after recalling for pedagogical purposes the demonstration of the particular Bjorken and Uraltsev sum rules at $w=1$ by this method \cite{LOPR-2,LOPR-3}, we perform a completely explicit demonstration of another important sum rule that involves only heavy mesons with $j^P = {1 \over 2}^-$ and their radial excitations. This latter sum rule can be phenomenologically useful because it constrains the derivatives of the radially excited Isgur-Wise functions at zero recoil. This demonstration, which requires a large set of calculations, is one of the of main objects of the rest of the paper.\par 

\subsection{Comments on other relativistic quark models}

Concerning the BSW quark model \cite{BSW}, we have demonstrated {\it numerically} \cite{LOR-6} that the IW function in this model yields, through eqn.(\ref{2.3e}), a measure that is indeed positive, which suggests also that this model is consistently satisfying the sum rule requirements. However, we did not provide an analytic demonstration of this feature. On the other hand, to our knowledge, excited states have not been studied in this model, and we do not have for the moment the possibility of testing directly the SR of the Bjorken-Uraltsev type for this model. 
Also, it is worth to point out that the BSW model exhibits a positive discrete $\delta$-function contribution to the measure (\ref{2.3e}) that cannot occur in the BT scheme.\par

As to other models, we do not claim to be exhaustive on the main problem of the present paper since we are quite aware that there are other relativistic quark models that are important in the literature. We cannot presently answer in general to the question as to whether they satisfy the sum rules.\par 

Among these other models, one can underline first the $P=\infty$ approach and the connected Melikhov dispersion relation approach to constituent relativistic quark model of form factors \cite{MELIKHOV-2} in meson decays and its heavy quark expansion  \cite{MELIKHOV-1}. We have argued that, because of its covariance in the heavy quark limit, the BT approach should be roughly speaking equivalent to the $P=\infty$ approach in this limit. Moreover, we have demonstrated (unpublished) that for a number of observables (IW functions, decay constants), both schemes give the same result.  Therefore, one expects them to satisfy the sum rules.

As to other relativistic quark models, we must emphasize :

(1) Faustov and Galkin relativistic quark model in the heavy quark limit, and their 
heavy quark $1/m_Q$ expansion of weak meson decay form factors \cite{FAUSTOV-GALKIN}. Within this scheme, a great variety of phenomena in heavy meson decays, in the heavy quark limit and at finite mass, have been studied. These studies involve the ground state heavy mesons, and also orbitally and radially excited states as well.

(2) Krutov, Shro and Troitsky relativistic quark model of constituent quarks \cite {KST}, in which properties of the elastic Isgur-Wise function have been investigated.

(3) Ivanov, Kalinovsky and Roberts model for heavy meson decays, based on the Dyson-Schwinger equation \cite{IKR}. 

Whether these models share the same good properties of the BT class is a problem beyond our present scope. It would require a serious, and explicit and detailed study. Moreover, the last theoretical scheme \cite{IKR}, being based on the Dyson-Schwinger equation, presents a different structure.\par 

\subsection{Outline of the present paper}

The rest of the paper is organized as follows. In Section 2 we recall the general OPE sum rules in heavy quark limit, and derive the relevant specific sum rules we are interested in here. In Section 3 we write down the explicit expressions for the BT model of the IW functions $\xi^{(n)}(w), \tau_{1/2}^{(n)}(w), \tau_{3/2}^{(n)}(w)$ corresponding respectively to the transitions ${1 \over 2}^- , \to {1 \over 2}^-, {1 \over 2}^+, {3 \over 2}^+$ and their radial excitations. In Section 4, we then demonstrate within the BT model the sum rules that we have chosen as examples. In Section 5 we point out weaknesses of the BT scheme, namely the lack of covariance of the form factors {\it at finite mass} and the related fact that sum rules with moments involving powers $(\Delta E^{(n)})^k$ with $k > 0$ do not hold in the model. Finally, in Section 6 we conclude. 

\section{Generalized Bjorken-Uraltsev sum rules}

Some years ago \cite{LOR-1, LOR-2} we did set a systematic method to obtain sum rules in the heavy quark limit of QCD, that relate the derivatives of the elastic Isgur-Wise function $\xi (w)$ to sums over inelastic IW functions to excited states. The method is based on the Operator Product Expansion (OPE) applied to heavy hadrons, and one of its key elements is the consideration, 
following Uraltsev \cite{URALTSEV}, of the non-forward amplitude, i.e. $B(v_i) \to D^{(n)}(v') \to B(v_f)$ with in general $v_i \not= v_f$. Then, the OPE side of the SR contains the elastic IW function $\xi (w_{if})$ and the SR depends in general on three variables, $w_i = v_i.v'$, $w_f = v_f.v'$ and $w_{if} = v_i.v_f$, that lie within a certain domain. By differentiation relatively to these variables within the domain and taking the limit to its boundary, one finds a very general class of SR that imply interesting consequences on the shape of $\xi (w)$. \par

To be more precise, as shown in \cite{LOR-1, LOR-2}, using the OPE as formulated for example in \cite{LMMOPR} and generalized to $v_i 
\not= v_f$ \cite{URALTSEV, LOR-1, LOR-2}, the trace formalism \cite{FALK} and arbitrary heavy quark currents
\beq
\label{1-1e}
J_1 = \bar{h}_{v'}^{(c)}\ \Gamma_1\ h_{v_i}^{(b)} \quad , \qquad J_2 
= \bar{h}_{v_f}^{(b)}\ \Gamma_2\ h_{v'}^{(c)}
\eeq

\noi the following sum rule can be written in the heavy quark limit \cite{LOR-1} :
\bea
\label{1-2e}
&&\Big \{ \sum_{D=P,V} \sum_n Tr \left [ \bar{\cal B}_f(v_f) 
\bar{\Gamma}_2 {\cal D}^{(n)}(v') \right ] Tr \left [ \bar{\cal 
D}^{(n)}(v') \Gamma_1 {\cal
B}_i(v_i)\right ] \xi^{(n)} (w_i) \xi^{(n)} (w_f) \nn \\
&&+ \ \hbox{Other excited states} \Big \}  = - 2 \xi (w_{if})Tr \left 
[ \bar{\cal B}_f(v_f) \bar{\Gamma}_2 P'_+ \Gamma_1 {\cal 
B}_i(v_i)\right ]
\eea

\noi In this formula $v'$ is the intermediate meson four-velocity, the projector
\beq
\label{1-3e}
P'_+ = {1 \over 2} (1 + {/ \hskip - 2 truemm v}')
\eeq

\noi comes from the residue of the positive energy part of the $c$-quark propagator, and $\xi (w_{if})$ is the elastic IW function that appears because one assumes a non-forward direction $v_i \not= v_f$. ${\cal B}_i$ and ${\cal B}_f$ are the $4 \times 4$ matrices of the ground state $B$ or $B^*$ meson and ${\cal D}^{(n)}$ those of all possible ground state or excited state $D$ mesons \cite{FALK} coupled to $B_i$ and $B_f$ through the currents. In formula (\ref{1-2e}) we have made explicit the $j = {1 \over 2}^-$ $D$ and $D^*$ mesons and their radial excitations, leaving implicit the sum over higher states. \par

The variables $w_i$, $w_f$ and $w_{if}$ are defined as \cite{LOR-1,LOR-2}
\beq \label{1-4e}
w_i = v_i \cdot v', \qquad w_f = v_f \cdot v', \qquad w_{if} = v_i \cdot v_f \ .
\eeq

\noi Their domain is
\beq
\label{1-5e}
w_i, w_f \geq 1\ , \quad w_i w_f - \sqrt{(w_i^2 - 1) (w_f^2 - 1)} 
\leq w_{if} \leq w_i w_f + \sqrt{(w_i^2 - 1) (w_f^2 - 1)}
\eeq

\noi and there is a subdomain for $w_i = w_f = w$ :
\beq
\label{1-6e}
w \geq 1, \qquad 1 \leq w_{if} \leq 2w^2-1 \ .
\eeq

Calling now $L(w_i, w_f, w_{if})$ the l.h.s. and $R(w_i, w_f, w_{if})$ the r.h.s. of (\ref{1-2e}), this SR writes
\beq
\label{1-7e}
L\left ( w_i, w_f, w_{if} \right ) = R \left ( w_i, w_f, w_{if} \right )
\eeq

\noi where $L(w_i, w_f, w_{if})$ is the sum over the intermediate $D$ states and $R(w_i, w_f, w_{if})$ is the OPE side. 
Within the domain (\ref{1-5e}) one can differentiate relatively to any of the variables $w_i$, $w_f$ and $w_{if}$
\beq
\label{1-8e}
{\partial^{p+q+r} L \over \partial w_i^p \partial w_f^q \partial 
w_{if}^r} = {\partial^{p+q+r} R \over \partial w_i^p \partial w_f^q 
\partial w_{if}^r}
\eeq

\noi and obtain different SR by taking limits to the frontiers of the domain.\par

Let us parametrize the elastic Isgur-Wise function $\xi (w)$ near zero recoil,
\beq
\label{1-9e}
\xi (w) = 1 - \rho^2 (w-1) + {\sigma^2 \over 2} (w-1)^2 - \cdots
\eeq

\noindent where $\xi'(1) = -\rho^2$ and $\xi''(1) = \sigma^2$ are the slope and the curvature.\par

From the SR (\ref{1-2e}), we gave in \cite{LOR-1, LOR-2} a simple and 
straightforward demonstration of Bjorken SR \cite{BJORKEN} and of another SR, that combined with the former, implied Uraltsev SR \cite{URALTSEV}.\par 
Bjorken and Uraltsev SR imply the lower bound on the elastic slope
\beq
\label{1-10e}
\rho^2 = - \xi ' (1) \geq {3 \over 4}
\eeq

\noi and the generalized SR imply the following lower bound on the curvature
\beq
\label{1-11e}
\sigma^2 = \xi '' (1) \geq {15 \over 16}
\eeq

A crucial simplifying feature of the calculation was to consider for the currents (\ref{1-1e}), vector or axial currents aligned along the 
initial and final velocities $v_i$ and $v_f$.\par 

In reference \cite{LOR-3}, exploiting a complete set of sum rules at a given order in the derivatives, we did obtain an improved new bound on the curvature
\beq
\label{1-12e}
\sigma^2 \geq {1 \over 5} \left[4\rho^2+3(\rho^2)^2\right]
\eeq

\noi that reduces to (\ref{1-11e}) for the lower bound for the slope (\ref{1-10e}).

\subsection{General sum rules}

As explained in detail in \cite{LOR-1, LOR-2, LOR-3}, if one uses in (\ref{1-1e},\ref{1-2e}) the vector currents
\beq
\label{10e}
J_1 = \bar{h}_{v'}^{(c)}\ {/ \hskip - 2 truemm v}_i\ h_{v_i}^{(b)}\ , 
\quad \qquad J_2 = \bar{h}_{v_f}^{(b)}\ {/ \hskip - 2 truemm v}_f\ 
h_{v'}^{(c)}
\eeq

\noi one obtains the so-called Vector Sum Rule
$$(w_i + 1) (w_f + 1) \sum_{\ell \geq 0} {\ell + 1 \over 2 \ell + 1} 
S_{\ell} (w_i, w_f, w_{if}) \sum_n \tau_{\ell + 1/2}^{(\ell)(n)}(w_i)
\tau_{\ell + 1/2}^{(\ell )(n)}(w_f)$$
\beq
\label{11e}
+\ \sum_{\ell \geq 1} S_{\ell} (w_i, w_f, w_{if}) \sum_n \tau_{\ell - 
1/2}^{(\ell)(n)}(w_i) \tau_{\ell - 1/2}^{(\ell )(n)}(w_f) = (1 + 
w_i+w_f+w_{if}) \xi(w_{if})
\eeq

\vskip 5 truemm
\noindent while chosing instead the axial currents
\beq
\label{12e}
J_1 = \bar{h}_{v'}^{(c)}\  {/ \hskip - 2 truemm v}_i \gamma_5 \ 
h_{v_i}^{(b)}\ , \quad \qquad J_2 = \bar{h}_{v_f}^{(b)}\ {/ \hskip - 2 
truemm v}_f
\gamma_5\ h_{v'}^{(c)}
\eeq

\noindent one finds the Axial Sum Rule
\bea
\label{13e}
&& \sum_{\ell \geq 0} S_{\ell + 1} (w_i, w_f, w_{if}) \sum_n 
\tau_{\ell + 1/2}^{(\ell)(n)}(w_i)
\tau_{\ell + 1/2}^{(\ell )(n)}(w_f)\nn \\
&&+ \ (w_i - 1) (w_f - 1) \sum_{\ell \geq 1} {\ell \over 2 \ell - 1} 
S_{\ell - 1} (w_i, w_f, w_{if}) \sum_n \tau_{\ell - 
1/2}^{(\ell)(n)}(w_i) \tau_{\ell -
1/2}^{(\ell )(n)}(w_f) \nn \\
&&= - (1 - w_i-w_f+w_{if}) \xi(w_{if}) 
\eea 

In the preceding expressions, following the formulation of heavy-light states for arbitrary $j^P$ 
given by Falk \cite{FALK}, we have defined in \cite{LOR-1, LOR-2, LOR-3} the IW 
functions $\tau_{\ell +
1/2}^{(\ell)(n)}(w)$ and $\tau_{\ell - 1/2}^{(\ell)(n)}(w)$, that 
correspond to the orbital angular momentum $\ell$ of the light quark 
relative to the heavy
quark, $j = \ell \pm {1 \over 2}$ being the total angular momentum of 
the light cloud, and $S_n(w_i,w_f,w_{if})$ is a Laguerre polynomial \cite{LOR-1}
\beq
\label{17e}
S_n(w_i,w_f,w_{if}) = \sum_{0 \leq k \leq {n \over 2}} C_{n,k} (w_i^2 
- 1)^k (w_f^2 - 1)^k (w_i w_f - w_{if})^{n-2k}
\eeq

\noindent with the coefficients
\beq
\label{18e}
C_{n,k} = (-1)^k {(n!)^2 \over (2n) !} \ {(2n - 2k) ! \over k! (n-k) 
! (n-2k)!} \ .
\eeq

\noi The precedent sums go over all the radial excitations, indicated by the index $n$.\par

\subsection{Bjorken and Uraltsev sum rules}

One obtains Bjorken SR from the Vector SR for $p = q = 0$ :
\beq
\label{28new}
\rho^2 = {1 \over 4} + {2 \over 3} \sum_{n\geq0} \mid\tau_{3/2}^{(1)(n)}(1)\mid^2 + {1 \over 4} \sum_n \mid
\tau_{1/2}^{(1)(n)}(1) \mid^2 \ .
\eeq

\noindent which, using the traditional notation \cite{IW-2}
\beq
\label{28bis}
\tau_{1/2}^{(n)}{(1)}(w) = 2 \tau_{1/2}^{(n)}(w) ,  \qquad \qquad \tau_{3/2}^{(1)(n)}(w) = \sqrt{3} \tau_{3/2}^{(n)}(w)
\eeq

\noindent writes 
\beq
\label{28new-bis}
\fbox{\ $\rho^2 = {1 \over 4} + 2 \sum_{n\geq0} \mid
\tau_{3/2}^{(n)}(1) \mid^2 + \sum_{n\geq0} \mid
\tau_{1/2}^{(n)}(1) \mid^2 $\  }
\eeq

From the Vector SR $p=2$, $q=0$ and $p=q=1$, or the Axial SR for $p = q = 0$ one gets 
\beq
\label{35new}
\rho^2 = \sum_{n\geq0} \mid\tau_{3/2}^{(1)(n)}(1) \mid^2
\eeq

\noindent Using Bjorken SR (\ref{28new}), relation (\ref{35new}) implies Uraltsev SR
\beq
\label{37new}
{1 \over 3} \sum_{n\geq0} \mid\tau_{3/2}^{(1)(n)}(1)\mid^2 - {1 
\over 4} \sum_{n\geq0} \mid\tau_{1/2}^{(1)(n)}(1)\mid^2 = {1 \over 
4}
\eeq

\noindent that, using the notation (\ref{28bis}) writes \cite{URALTSEV}
\beq
\label{37new-bis}
\fbox{\ $\sum_{n\geq0} \mid\tau_{3/2}^{(n)}(1)\mid^2 - \sum_{n\geq0} \mid\tau_{1/2}^{(n)}(1)\mid^2 = {1 \over 
4}$ \ }
\eeq

\subsection{Sum rule involving only IW functions of $j^P = {1 \over 2}^-$ heavy mesons and their radial excitations}

From (\ref{1-8e}), differentiating the Vector SR if $p+q=2$ and the Axial SR if $p+q=3$ one finds a whole set of rather involved SR, that reduce to the following linearly independent relations \cite{LOR-3} :
\beq
\label{54new}
\rho^2 = - \ {4 \over 5} \sum_{n\geq0} \tau_{3/2}^{(1)(n)}(1) 
\tau_{3/2}^{(1)(n)'}(1) + {3 \over 5} \sum_{n\geq0} \tau_{1/2}^{(1)(n)}(1) 
\tau_{1/2}^{(1)(n)'}(1)
\eeq
\beq
\label{55new}
\sigma^2  = - \sum_{n\geq0} \tau_{3/2}^{(1)(n)}(1) \tau_{3/2}^{(1)(n)'}(1)
\eeq
\beq
\label{56new}
\sigma^2  = 2 \sum_{n\geq0} \mid\tau_{5/2}^{(2)(n)}(1)\mid^2
\eeq
\beq
\label{57new}
\rho^2  - {4 \over 5} \sigma^2 + \sum_{n\geq0} \mid
\tau_{3/2}^{(2)(n)}(1) \mid^2 = 0
\eeq
\beq
\label{58new}
\fbox{\ ${4 \over 3} \rho^2  - {5 \over 3} \sigma^2 + \sum_{n\geq0} \mid\xi^{(n)'}(1) \mid^2 = 0$\ }
\eeq

\noi The last relation (\ref{58new}) depends on quantities involving only ${1 \over 2}^-$ states, namely the slope $\rho^2$ and curvature $\sigma^2$ of the elastic IW function $\xi(w)$ and the sum $\sum_{n\geq0} \left [\xi^{(n)'}(1) \right ]^2$ depends on the derivatives of the IW to all the ${1 \over 2}^-$ states, where $n = 0$ corresponds to the ground state, and $n \not= 0$ to its radially excited states.  Equation (\ref{58new}) was the main result obtained in \cite{LOR-3} that, isolating the ground state, can be written in the form 
\beq
\label{59new}
{4 \over 3} \rho^2 + (\rho^2)^2 - {5 \over 3} \sigma^2 + \sum_{n\geq1} \mid\xi^{(n)'}(1) \mid^2 = 0
\eeq

\noi This last relation implies the improved lower bound (\ref{1-12e}) on the curvature.\par 

\section{Bakamjian-Thomas relativistic quark models}

The Bakamjian-Thomas relativistic scheme \cite{BT,TERENTEV,KEISTER,CARDARELLI,LOPR-1} is a class of models with a fixed number of constituents in which the states are covariant under the Poincar\'e group. The model relies on an appropriate Lorentz boost of the eigenfunctions of a Mass Operator or Hamiltonian describing the hadron spectrum at rest.\par

Unfortunately, the matrix elements of the usual {\it one-body additive} current operators are not covariant. 
However, for the meson ground state \cite{LOPR-1} we found the important feature that, in the heavy quark limit, the current matrix elements, {\it when the current is coupled to the heavy quark}, are Lorentz covariant. Therefore, the IW function can be computed without any ambiguity. We have extended this result to the matrix elements between the ground sate and P-wave excited states \cite{MORENAS-1,MORENAS-2}.\par

Moreover, these matrix elements in the heavy quark limit exhibit Isgur-Wise scaling \cite{IW-1}. Given a Mass Operator $M$ describing the spectrum at rest, with the only constraint of being rotationally invariant, the model provides an unambiguous result for the Isgur-Wise functions, e.g. in particular the elastic $\xi (w)$ \cite{LOPR-1} and the inelastic to P-wave states $\tau_{1/2}(w)$, $\tau_{3/2}(w)$ \cite{MORENAS-1}.\par 

On the other hand, the sum rules (SR) in the heavy quark limit of QCD, like Bjorken and Uraltsev SR are analytically satisfied in the model \cite{LOPR-2,LOPR-3}.\par
  
The BT framework is a class of relativistic quark models, since there is a great arbitrariness in the Mass Operator $M$. In \cite{MORENAS-2}, we have chosen the Godfrey-Isgur Hamitonian \cite{GI}, that gives a very complete description of the light $q\overline{q}$ and heavy $Q\overline{q}$ meson spectra in order {\it to predict} within the BT scheme the corresponding IW functions for the ground state and the excited states.\par

\subsection{Isgur-Wise functions within the BT model}

Let us use the following notation : the three-momentum of the light quark is $\vec p$ and its energy $p^0 = \sqrt{m^2+p^2}$, with $p =\ \mid {\vec p} \mid$. The expression of the ${1 \over 2}^- \to {1 \over 2}^-$ IW functions in the BT model is given by the integral \cite{LOPR-1}
$$\xi^{(n)}(v.v') = {1 \over 1+v.v'} \int {d{\vec p} \over (2\pi)^3}\ {\sqrt{(p.v)(p.v')} \over p^0}\ {m(v.v'+1)+p.(v+v') \over \sqrt{(p.v+m)(p.v'+m)}}$$
\beq
\label{3-1e}
\times\ \varphi^{(n)}\left(\sqrt{(p.v')^2-m^2}\right)^* \varphi\left(\sqrt{(p.v)^2-m^2}\right) 
\eeq

\noindent where the superscript $(n)$ labels the radial excitations. The elastic IW function corresponds to $\xi(w) = \xi^{(0)}(w)$, with the wave function $\varphi = \varphi^{(0)}$.\par
On the other hand, the IW functions for the transitions ${1 \over 2}^- \to {1 \over 2}^+$ and ${1 \over 2}^- \to {3 \over 2}^+$ are respectively given by \cite{MORENAS-1,MORENAS-2}
$$\tau_{1/2}^{(n)}(w) = - {1 \over 2(w-1)}\ \int {d^3\vec p \over p^0}\ \varphi^{(n)}_{1/2}(\sqrt{(p.v')^2-m^2})^*\ \varphi(\sqrt{(p.v)^2-m^2})$$
\beq
\label{105-74e}
\times\ {1 \over \sqrt{(p.v+m)(p.v'+m)}}\ {1 \over \sqrt{(p.v')^2-m^2}}\ [(p.v')+m][(p.v')-(p.v)+m(w-1)]  
\eeq

$$\tau_{3/2}^{(n)}(w) = - {1 \over 2(w-1)(w+1)^2}\ \int {d^3\vec p \over p^0}\ \varphi^{(n)}_{3/2}(\sqrt{(p.v')^2-m^2})^*\ \varphi(\sqrt{(p.v)^2-m^2})$$
$$\times\ {1 \over \sqrt{(p.v+m)(p.v'+m)}}\ {1 \over  \sqrt{(p.v')^2-m^2}}\ [-3(p.v)^2+(2w-1)(p.v')^2$$
\beq
\label{105-84e}
+\ 2(2w-1)(p.v)(p.v')+2(w+1)(w(p.v')-(p.v))m-(w^2-1)m^2] 
\eeq

The radial wave functions for the ${1 \over 2}^-$ states $\varphi^{(n)}(|{\vec p}|)$ and for the orbitally excited states ${1 \over 2}^+$ and ${3 \over 2}^+$, $\varphi^{(n)}_{1/2}(|{\vec p}|)$ and $\varphi^{(n)}_{3/2}(|{\vec p}|)$, are normalized according to
\beq
\label{3-2e}
\int {d{\vec p} \over (2\pi)^3}\ |\varphi(|{\vec p}|)|^2 = 1 
\eeq

\noindent Notice that for the P-wave states this normalization is different from the ones used in refs. \cite{MORENAS-1,MORENAS-2,LOR-6}.

\section{Explicit proof of some important sum rules within the Bakamjian-Thomas scheme}

In this Section we demonstrate very explicitly a number of SR within the BT scheme, namely (\ref{28new-bis}), (\ref{37new-bis}) and (\ref{58new}), just to consider some important examples. To this aim, we will deduce rather cumbersome expressions. We think that it is worth to write down these formulas in order to illustrate the fact that a powerful theorem, based on the Lorentz group representation underlying the BT scheme, is at the basis of the satisfaction of the SR in the BT approach.\par

Indeed, switching off hard gluon radiative corrections, we have assumed in Lorentz I \cite{LOR-5} and Lorentz II \cite{LOR-6} that a current matrix element factorizes, in general and also within the BT scheme, into a heavy quark part and a light cloud overlap. Completeness in this Hilbert space implies the Bjorken-Uraltsev sum rules. In Section 11 of \cite{LOR-6} we have described the Lorentz group representation that acts on the light cloud Hilbert space within the Bakamjian-Thomas framework.\par

Of course, it would be a further very direct check of these results to start from the current matrix element within the BT model and to demonstrate that it factorizes into a heavy quark current matrix element and a light cloud overlap, and that the result is covariant in the heavy quark limit. This will be the object of further investigation.

\subsection{Relevant quantities at zero recoil}

With the IW functions in the BT framework (\ref{3-1e}), (\ref{105-74e}) and (\ref{105-84e}) at hand, we are now in the position of verifying that the BT scheme explicitly satisifies Bjorken SR (\ref{28new-bis}), Uraltsev SR (\ref{37new-bis}) and also the relation (\ref{58new}), that involves only ${1 \over 2}^-$ states. These relations depend on the IW functions (\ref{3-1e}), (\ref{105-74e}) and (\ref{105-84e}), while (\ref{56new}) and (\ref{57new}) involve $\ell = 2$ IW functions, not given here explicitly in the BT model. Of course, it is immediate to demonstrate the other SR (\ref{54new}) and (\ref{55new}) as well.

The IW functions of the BT model are covariant, as we can see by inspection of the particular cases (\ref{3-1e}), (\ref{105-74e}) and (\ref{105-84e}), and therefore we can make the calculations in any reference frame. For our purposes, we chose the following frame :
\beq
\label{3-3e}
v = (w,0,0,-\sqrt{w^2-1})\ , \qquad \qquad \qquad v' = (1,0,0,0) 
\eeq

\noindent that will be convenient to compute the sum over radial excitations in the sums over $n$ in (\ref{28new-bis}), (\ref{37new-bis}) and (\ref{58new}).\par
Let us first consider the elastic IW function $\xi(w)$. Performing an expansion of (\ref{3-1e}) in powers of $w-1$, one finds of course $\xi(1) = 1$ and the following expressions for the slope
$$\xi'(1) = {1 \over 24\pi^2}\ \int_0^\infty dp\ \varphi(p)^* {p \over (m^2+p^2)(m+\sqrt{m^2+p^2})}$$
$$\times\ \{mp (5m^2+4p^2+m\sqrt{m^2+p^2}) \varphi(p)$$
\beq
\label{3-4e}
+\ 4(m^2+p^2)(m+\sqrt{m^2+p^2})\ [2(m^2+2p^2) \varphi'(p) + p(m^2+p^2) \varphi''(p)]\}
\eeq

\noindent and the curvature
$$\xi''(1) = - {1 \over 480\pi^2} \int_0^\infty dp\ \varphi(p)^* {p \over (m^2+p^2)^2(m+\sqrt{m^2+p^2})}$$
$$\times\ \{mp (127m^4+208m^2p^2+96p^4+23m^3\sqrt{m^2+p^2}+8mp^2\sqrt{m^2+p^2})\ \varphi(p)$$
$$+\ 8(m^2+p^2)^2 [2m(m^2-6p^2+m\sqrt{m^2+p^2})\ \varphi'(p)$$
$$-\ p(49m^3+64mp^2+45m^2\sqrt{m^2+p^2}+60p^2\sqrt{m^2+p^2})\ \varphi''(p)$$
\beq
\label{3-5e}
-\ 2(m^2+p^2)(m+\sqrt{m^2+p^2})(4(m^2+3p^2)\ \varphi^{(3)}(p) + p(m^2+p^2)\ \varphi^{(4)}(p))] \}
\eeq

For the transitions to the positive parity excited states we find, at zero recoil, from (\ref{105-74e}) and (\ref{105-84e}) :

$$\tau_{1/2}^{(n)}(1) = {1 \over 12 \pi^2} \int_0^\infty dp\ \varphi^{(n)}_{1/2}(p)^* {p \over \sqrt{p^2+m^2}}$$
\beq
\label{4-2e}
\times\ \left[(2m^2+3p^2-2m\sqrt{p^2+m^2}) \varphi(p) + 2p(p^2+m^2) \varphi'(p) \right]
\eeq

$$\tau_{3/2}^{(n)}(1) = {1 \over 12 \pi^2} \int_0^\infty dp\ \varphi_{3/2}^{(n)}(p)^* {p^2 \over \sqrt{p^2+m^2}(m+\sqrt{p^2+m^2})}$$
\beq
\label{4-3e}
\times\ \left[mp \varphi(p) + 2(p^2+m^2)(m+\sqrt{p^2+m^2}) \varphi'(p) \right]
\eeq

In the preceding expressions $\varphi(p)$, $\varphi^{(n)}_{1/2}(p)$ and $\varphi^{(n)}_{3/2}(p)$ are the radial wave functions for the states ${1 \over 2}^-, {1 \over 2}^+$ and ${3 \over 2}^+$, normalized according to (\ref{3-2e}), or
\beq
\label{3-6e}
{1 \over 2\pi^2} \int_0^\infty p^2 dp |\varphi(p)|^2 = {1 \over 2\pi^2} \int_0^\infty p^2 dp |\varphi^{(n)}_j(p)|^2  = 1 \qquad \left(j = {1 \over 2}\ , {3 \over 2}\right)
\eeq

In consistency with this normalization, the completeness relation reads, in each of the considered sectors :
\beq
\label{3-7e}
\sum_{n\geq 0} \varphi^{(n)}(p') \varphi^{(n)}(p)^* = \sum_{n\geq 0} \varphi^{(n)}_j(p') \varphi^{(n)}_j(p)^* = 2\pi^2\ {\delta(p-p') \over p^2} \qquad \left(j = {1 \over 2}\ , {3 \over 2}\right)
\eeq

\subsection{Bjorken and Uraltsev sum rules in the BT model}

Let us now compute the sums over $n$ in expressions (\ref{28new-bis}) and (\ref{37new-bis}) :
$$\sum_{n\geq 0}\ \mid \tau_{1/2}^{(n)}(1) \mid^2\ = \sum_{n\geq 0} \left( {1 \over 12\pi^2} \right)^2\ \int_0^\infty dp' \ \int_0^\infty dp$$
$$\left[(2m^2+3p'^2-2m\sqrt{p'^2+m^2}) \varphi(p')^* + 2p'(p'^2+m^2) \varphi'(p')^* \right]$$
$${p' \over \sqrt{p'^2+m^2}}\ \left(\sum_{n\geq 0} \varphi^{(n)}_{1/2}(p') \varphi^{(n)}_{1/2}(p)^* \right)\ {p \over \sqrt{p^2+m^2}}$$
\beq
\label{4-4e}
\left[(2m^2+3p^2-2m\sqrt{p^2+m^2}) \varphi(p) + 2p(p^2+m^2) \varphi'(p) \right]
\eeq

$$\sum_{n\geq 0}\ \mid \tau_{3/2}^{(n)}(1) \mid^2\ = \sum_{n\geq 0} \left( {1 \over 12\pi^2} \right)^2\ \int_0^\infty dp' \ \int_0^\infty dp$$
$$\left[mp' \varphi(p')^* + 2(p'^2+m^2)(m+\sqrt{p'^2+m^2}) \varphi'(p')^* \right]$$
$${p'^2 \over \sqrt{p'^2+m^2}(m+\sqrt{p'^2+m^2})}\ \left(\sum_{n\geq 0} \varphi^{(n)}_{3/2}(p') \varphi^{(n)}_{3/2}(p)^* \right)\ {p^2 \over \sqrt{p^2+m^2}(m+\sqrt{p^2+m^2})}$$
\beq
\label{4-5e}
\left[mp \varphi(p) + 2(p^2+m^2)(m+\sqrt{p^2+m^2}) \varphi'(p) \right]
\eeq

Using the completeness relation (\ref{3-6e}), the combination that appears in the Bjorken sum rule writes then, 
$$2\sum_{n\geq 0}\ \mid \tau_{3/2}^{(n)}(1) \mid^2+\sum_{n\geq 0}\ \mid \tau_{1/2}^{(n)}(1) \mid^2\ 
= {1 \over 24\pi^2} \int_0^\infty dp\ {p^2 \over (p^2+m^2)(m+\sqrt{p^2+m^2})^2}$$
$$\times \{\varphi(p)^*[(p^2(4m^2+3p^2+2m\sqrt{p^2+m^2})\varphi(p)+2p(p^2+m^2)(p^2+2m^2+2m\sqrt{p^2+m^2})\varphi'(p)]$$
\beq
\label{4-6e}
+\ \varphi'(p)^* 2\sqrt{p^2+m^2}[(p^2+2m^2)(m+\sqrt{p^2+m^2})(p \varphi(p)+2(p^2+m^2)\varphi'(p))]\}
\eeq

\noi To check Bjorken sum rule, we need to compare (\ref{4-6e}) with $-\xi'(1) - {1 \over 4} = \rho^2 - {1 \over 4}$, where $- \rho^2$ is given by (\ref{3-4e}). Since in the latter expression appears the second derivative $\varphi''(p)$, to make the comparison we need to integrate by parts the term proportional to $\varphi'(p)^*$ in (\ref{4-6e}). After this operation is done, we find, taking into account the normalization (\ref{3-6e}) to transform accordingly the numerical contribution ${1 \over 4}$ :
$${1 \over 4} + 2\sum_{n\geq 0}\ \mid \tau_{3/2}^{(n)}(1) \mid^2+\sum_{n\geq 0}\ \mid \tau_{1/2}^{(n)}(1) \mid^2\ $$ 
\beq
\label{4-8e}
= - {1 \over 24\pi^2}\ \int_0^\infty dp\ \varphi(p)^* {p \over (m^2+p^2)(m+\sqrt{m^2+p^2})} 
\eeq
$$\times\ \{mp (5m^2+4p^2+m\sqrt{m^2+p^2}) \varphi(p)$$

\noindent i.e., we find the expression for $- \xi'(1)$ (\ref{3-4e}), and Bjorken sum rule is demonstrated. 

The relevant expression for the Uraltsev sum rule is given by
$$\sum_{n\geq 0}\ \mid \tau_{3/2}^{(n)}(1) \mid^2-\sum_{n\geq 0}\ \mid \tau_{1/2}^{(n)}(1) \mid^2\ = - {1 \over 24\pi^2} \int_0^\infty dp\ {p^3 \over \sqrt{p^2+m^2}(m+\sqrt{p^2+m^2})^2}$$
$$\times \{\varphi(p)^*[p(2m+3\sqrt{p^2+m^2})\varphi(p)+2(p^2+m^2)(m+\sqrt{p^2+m^2})\varphi'(p)]$$
\beq
\label{4-7e}
+\ \varphi'(p)^* [2(p^2+m^2)(m+\sqrt{p^2+m^2})] \varphi(p)\}
\eeq

Integrating again by parts the term proportional to $\varphi'(p)^*$, and taking into account the normalization (\ref{3-6e}), we find that Uraltsev sum rule (\ref{37new-bis}) is also satisfied in the BT model.
 
\subsection{Sum rule involving radially excitated $j^P = {1 \over 2}^-$ heavy mesons in the BT model}

 Let us now compute the sum over the radial excitations appearing in the SR (\ref{58new}). Using expression (\ref{3-4e}) with the necessary replacement for the $n$-th radial excitation $\varphi(p)^* \to \varphi^{(n)}(p)^*$, we have to deal with the expression
$$\sum_{n\geq 0}\ \mid\xi^{(n)'}(1)\mid^2\ = \sum_{n\geq 0} \left( {1 \over 24\pi^2} \right)^2\ \int_0^\infty dp' \ \int_0^\infty dp$$
$$\{mp' (5m^2+4p'^2+m\sqrt{m^2+p'^2}) \varphi(p')^*$$
$$+\ 4(m^2+p'^2)(m+\sqrt{m^2+p'^2})\ [2(m^2+2p'^2) \varphi'(p')^* + p'(m^2+p'^2) \varphi''(p')^*]\}$$
$${p' \over (m^2+p'^2)(m+\sqrt{m^2+p'^2})}\ \left(\sum_{n\geq 0} \varphi^{(n)}(p') \varphi^{(n)}(p)^* \right)\ {p \over (m^2+p^2)(m+\sqrt{m^2+p^2})}$$
\beq
\label{3-7-1e}
\{mp (5m^2+4p^2+m\sqrt{m^2+p^2}) \varphi(p)
\eeq
$$+\ 4(m^2+p^2)(m+\sqrt{m^2+p^2})\ [2(m^2+2p^2) \varphi'(p) + p(m^2+p^2) \varphi''(p)]\}$$

\noindent Using now the completeness relation for the radial wave functions (\ref{3-6e}), one finds
$$\sum_{n\geq 0} \mid\xi^{(n)'}(1)\mid^2\ = \sum_{n\geq 0}\ {1 \over 288\pi^2}\ \int_0^\infty dp\ {1 \over (m^2+p^2)^2(m+\sqrt{m^2+p^2})^2}$$
$$\{mp (5m^2+4p^2+m\sqrt{m^2+p^2}) \varphi(p)^*$$
\beq
\label{3-7-2e}
+\ 4(m^2+p^2)(m+\sqrt{m^2+p^2})\ [2(m^2+2p^2) \varphi'(p)^* + p(m^2+p^2) \varphi''(p)^*]\}
\eeq
$$\{mp (5m^2+4p^2+m\sqrt{m^2+p^2}) \varphi(p)$$
$$+\ 4(m^2+p^2)(m+\sqrt{m^2+p^2})\ [2(m^2+2p^2) \varphi'(p) + p(m^2+p^2) \varphi''(p)]\}$$

To be able to compare (\ref{3-7-2e}) with the expressions (\ref{3-4e}) and (\ref{3-5e}) for the slope and the curvature, we need to integrate twice by parts the precedent formula. Doing this, one finds
$$\sum_{n\geq 0} \mid\xi^{(n)'}(1)\mid^2\ = {1 \over 288\pi^2}\ \int_0^\infty dp\ \varphi(p)^*\ {p \over (m^2+p^2)^2(m+\sqrt{m^2+p^2})^2}$$
$$\times\ \{-m p (15m^4 + 15m^3\sqrt{m^2+p^2} - 8m(m^2+p^2)^{3/2}+32(m^2+p^2)^2)\ \varphi(p)$$
\beq
\label{3-8e}
+\ 8(m^2+p^2)^2 [(-30m^3-18m^2\sqrt{m^2+p^2}+44m(m^2+p^2)+32(m^2+p^2)^{3/2})\ \varphi'(p)
\eeq
$$+\ p (-15m^3-15m^2\sqrt{m^2+p^2}+72m(m^2+p^2)+68(m^2+p^2)^{3/2})\ \varphi''(p)$$
$$+\ 2(m^2+p^2)(m+\sqrt{m^2+p^2}) ((-8m^2+12(m^2+p^2))\varphi^{(3)}(p)+p(m^2+p^2)\ \varphi^{(4)}(p))]\}$$

Gathering formulas (\ref{3-4e}), (\ref{3-5e}) and (\ref{3-8e}) one can check that the sum rule (\ref{58new}) is satisfied in the BT model.\par

The SR (\ref{59new}) can be phenomenologically useful since it gives an upper bound on the derivative at zero recoil of any inelastic IW function between the ground state and a radial excitation $\xi^{(n)}(w)$, that vanishes at zero recoil $\xi^{(n)}(1) = 0\ (n \not= 0)$ : 
\beq
\label{3-9e}
\mid\xi^{(n)'}(1) \mid^2\ \leq\  {5 \over 3} \sigma^2 - \left[ {4 \over 3} \rho^2 + (\rho^2)^2\right] \qquad \qquad (n \not= 0)
\eeq

The r.h.s. of the precedent inequality is positive because of the lower bound on the curvature (\ref{1-12e}). The physics involved in this inequality will be examined elsewhere.

\section{Problems of the Bakamjian-Thomas scheme \\ for current matrix elements at finite mass}

The Bakamjian-Thomas (BT) relativistic quark models \cite{BT,TERENTEV,KEISTER,CARDARELLI,LOPR-1} are a class of models with a fixed number of constituents in which the states are covariant under the Poincar\'e group. The model relies on an appropriate Lorentz boost of the eigenfunctions of a Hamiltonian describing the hadron spectrum at rest.\par 
We have proposed a formulation of this scheme for the meson ground states \cite{LOPR-1} and demonstrated the important feature that, in the heavy quark limit, the current matrix elements, when the current is coupled to the heavy quark, are {\it covariant}. We have extended this scheme to P-wave excited states \cite{MORENAS-1}.\par

As pointed out above, these matrix elements in the heavy quark limit exhibit Isgur-Wise (IW) scaling. As demonstrated in \cite{LOPR-1, MORENAS-1}, given a Hamiltonian describing the spectrum, the model provides an unambiguous result for the Isgur-Wise functions, the elastic $\xi (w)$ \cite{IW-1} and the inelastic to P-wave states $\tau_{1/2}(w)$, $\tau_{3/2}(w)$.\par 

However, these interesting and encouraging results do not hold when including the $1/m_Q$ corrections in the heavy quark mass expansion. Indeed, the heavy meson current matrix elements $< \vec{P'}, \epsilon' | J | \vec{P}, \epsilon >$, where the current $J = \overline{Q}' \Gamma Q$ {\it acts on the heavy quark}, are only covariant in the heavy quark limit. 
Moreover, in the BT scheme, current conservation only holds in the heavy quark limit, and turns out to be violated at finite mass.\par

The higher moment SR of the heavy quark limit of QCD involve powers of level spacings $\Delta E^{(n)}$, i.e. sums for example of the form $\sum_n (\Delta E^{(n)}_j)^{k} \mid \tau_j^{(n)}(w) \mid^2$ ($j = {1 \over 2},  {3 \over 2}$) where $k > 0$ is an integer \cite{VOLOSHIN,LMMOPR,URALTSEV,BIGI,GK}, follow from the application of the OPE  to the power corrections $1/m_Q^k$, i.e. the identification of the sum over intermediate heavy hadrons with the corresponding short distance counterpart beyond the heavy quark limit of QCD. But since the BT scheme is not covariant at finite mass, there is no hope of obtaining these SR.

We have extended our study of form factors in the BT approach to finite mass in a recent work \cite{DLOR}, studying the ground state decays ${1 \over 2}^- \to {1 \over 2}^-$, namely $\overline{B} \to D(D^*) \ell \nu$, and the decays ${1 \over 2}^- \to {1 \over 2}^+, {3 \over 2}^+$ to the positive parity excited states $\overline{B} \to D^{**} \ell \nu$, and have exposed in detail the successes and the problems that one encounters.

\section{Conclusions}

The aim of this paper is well defined and relatively simple, namely to show explicitly that the Bakamjian-Thomas scheme satisfies the sum rules of the Bjorken-Uraltsev type of the heavy quark limit of QCD (lowest moment SR). 

In this paper we have illustrated this general statement by the explicit demonstration of several physically significant sum rules, namely Bjorken and Uraltsev SR and another one that involves only the ${1 \over 2}^-$ states and their radial excitations. In a similar way, one could demonstrate within the BT approach any of the SR of the heavy quark limit of QCD.\par 

The SR involving transitions to radially excited states is interesting, as it gives an upper bound on the derivative at zero recoil $\mid \xi^{(n)'}(1) \mid$ of any inelastic IW function between the ground state and a radial excitation - the IW function itself vanishes at zero recoil $\xi^{(n)}(1) = 0$ for $n \not= 0$. At the same time, for a given ground state IW function, it leads to a useful bound on the magnitude of curvature. \par

One must finally keep in mind that what precedes and all the construction described in this paper, are only valid in a particular approximation, namely if one neglects the hard gluon radiative corrections, as we have emphasized in previous work \cite{LOR-5,LOR-6}.

Let us recall that taking into account hard gluons or UV divergencies, the overlaps over light quark states and their sums (given by HQET) become $\mu$ dependent, with the matching Wilson coefficients $C_i(\mu)$ cancelling this dependence for physical matrix elements of currents \cite{URALTSEV-RAD,DORSTEN}.

On the other hand, we have recalled the problems of the BT scheme at finite mass due to the fact that the covariance of form factors, with the current coupled to the heavy quark, holds only in the heavy quark limit.

\section*{Postface} \hspace*{\parindent} 
On the 6th October 2015, Jean-Claude Raynal suddenly passed away. Jean-Claude was a brilliant physicist and also a very gifted mathematician. He has been our very dear friend and collaborator for many years, and we are overwhelmed by this awful and unexpected event. We have payed hommage to his scientific work on January 18th 2016 at the Laboratoire de Physique Th\'eorique of Orsay.  

\end{document}